\documentclass[a4paper,10pt,twoside]{cpc-hepnp}

\usepackage{multicol}
\usepackage{graphicx}
\usepackage{booktabs}
\usepackage{amssymb,bm,mathrsfs,bbm,amscd}
\usepackage[tbtags]{amsmath}
\usepackage{lastpage}

\usepackage{color}
\usepackage{natbib}

\begin{document}
%\begin{CJK}{GBK}{kai}

\fancyhead[co]{\footnotesize ZHANG Tong~~et~al: Proposal for High-harmonic EEHG Lasing at Shanghai Deep Ultra-Violet Free-electron Laser}

\footnotetext[0]{Received July 2, 2013}

\title{Proposal for High-harmonic EEHG Lasing at Shanghai Deep Ultra-Violet Free-electron Laser~\thanks{Supported by Major State Basic Research Development Program of China (2011CB808300), and Natural Science Foundation of China (11075199 and 11175240)}}

\author{%
	  ZHANG~Tong$^{1}$%\email{zhangtong@sinap.ac.cn}%
\quad FENG~Chao$^{1}$
\quad DENG~Hai-Xiao$^{1}$\\
\quad WANG~Dong$^{1;1)}$\email{wangdong@sinap.ac.cn}%
\quad ZHAO~Zhen-Tang$^{1}$
}

\maketitle

\address{%
$^1$ Shanghai Institute of Applied Physics, Chinese Academy of Sciences, Shanghai 201800, China
%$^2$ University of Chinese Academy of Sciences, Beijing, 100049, China
}

\begin{abstract}
  The echo-enabled harmonic generation (EEHG) free-electron laser (FEL) has been already demonstrated at lower harmonics and the first lasing at third harmonic also has been achieved at Shanghai deep ultra-violet FEL (SDUV-FEL). While the great advantage of much higher harmonic up-conversion efficiency of EEHG over other seeded FELs only shows evidently at much higher harmonics. In this paper, we investigate the possibility of EEHG lasing at 10-th harmonic of the seed laser at SDUV-FEL, both physical designs and numerical simulations have been studied carefully. Two proposals of EEHG at 10-th harmonic have been studied respectively, i.e. with the seed lasers of the same color and two difference colors, the simulation results indicate that both approaches could be the candidate for EEHG lasing at 10-th harmonic at SDUV-FEL, meanwhile the coherent synchrotron radiation does not affect the performance of EEHG-FEL but only slightly shifts the central radiation frequency.
\end{abstract}



\begin{multicols}{2}

\section{Introduction}
The free-electron lasers (FELs) have been proved to be the novel advanced coherent synchrotron radiation light source~\cite{Madey_1971_FEL_JAP,OShea_2001_FEL_Science,Barletta_2010_FEL_NIMA,McNeil_2010_XFEL_NP}, especially with the commissioning of the world¡¯s first two hard X-ray FELs, linear coherent light source (LCLS)~\cite{Emma_2010_LCLS_NP} and Spring-8 {\AA}ngstr{\"o}m compact free-electron laser (SACLA)~\cite{Ishikawa_2012_SACLA_NP}. However, all the hard X-ray FELs together with the next coming European XFEL~\cite{Massimo_2007_CDREXFEL,Altarelli_2011_EXFEL_NIMB} and SwissFEL~\cite{SwissFEL_CDR}, etc. are all working on the self-amplified spontaneous emission (SASE) principle~\cite{Bonifacio_1984_FEL_OC,Milton_2001_SASE_Science}, thus large intensity fluctuation and poor temporal coherency are inevitable. In order to obtain the truly full coherent FEL radiations, much effects have been put to extend the FEL working schemes, among which the external optical laser seeded FEL principles have been regarded as the key approaches to lead the FEL radiations to be fully coherent both spatially and temporally~\cite{Yu_2000_HGHG_Science,Jia_2008_EHGHG_APL,Stupakov_2009_EEHG_PRL,Xiang_2010_EEHG_PRL,Zhao_2012_SDUVEEHG_NP}.

Experimentally, fully coherent hard X-ray FELs from self-seeding approaches have been demonstrated at LCLS for the first time~\cite{Amann_2012_LCLSself-seeding_NP}, in which the generated SASE from the first undulator stage continue seeds the time-shifted electron bunch to produce FELs with much purer spectra and much higher power intensity. While the external seeded FELs now have been pushed into EUV regime at FERMI@Elettra FEL in which the high-gain harmonic generation (HGHG) principle is utilized~\cite{Allaria_2012_FermiFEL_NP}. Other HGHG FELs worldwide (e.g. ATF@BNL~\cite{Doyuran_2001_BNL_PRL} and SDUV-FEL@SINAP~\cite{Zhao_2004_SDUV_NIMA}) indicate that powerful FEL pulses with almost Fourier transform limited could be achieved. Furthermore, another seeded FEL principle, so-called echo-enabled harmonic generation (EEHG), known as the much higher up-frequency conversion efficiency was originally proposed by Dr.~Stupakov at 2009~\cite{Stupakov_2009_EEHG_PRL}. In the next two years, demonstration experiments at NLCTA~\cite{Xiang_2010_EEHG_PRL} and SDUV-FEL~\cite{Zhao_2010_SDUV-FEL_FEL10} have been successfully performed and the first lasing of EEHG at third harmonic has already been successfully achieved at SDUV-FEL~\cite{Zhao_2012_SDUVEEHG_NP}, which shows the great potential of EEHG FELs. Up to now, several planned or running FEL facilities, e.g. FLASH-II~\cite{Feldhaus_2010_FLASH_JPB}, NGLS~\cite{Corlett_2011_NGLS_PAC2011}, SXFEL~\cite{SXFEL_CDR}, SwissFEL~\cite{SwissFEL_CDR} etc., take EEHG as one of the most important options to push the FEL wavelength into the X-ray regime.

In this paper, we discuss the possibility of much higher harmonic EEHG lasing at SDUV-FEL, i.e. EEHG at 10-th harmonic of the seed laser (EEHG-10), with the purpose of demonstrating the great advantage of EEHG over HGHG at much higher harmonics. Comprehensive studies including two EEHG-10 proposals are presented, i.e. EEHG with two seed lasers of the same frequency and different frequencies. Numerical simulations indicate that with minor modifications of the present hardware configuration at SDUV-FEL, EEHG could be lasing at 10-th harmonic while the HGHG signal is deeply suppressed.

\section{Overview of EEHG-10@SDUV-FEL}
Shanghai deep ultra-violet free-electron laser (SDUV-FEL) is a test facility for versatile novel FEL principles~\cite{Zhao_2004_SDUV_NIMA}. Up to now the FEL experiments including SASE~\cite{Dongguo_2010_SDUVSASE_FEL10}, HGHG~\cite{Zhao_2010_SDUV-FEL_FEL10} and EEHG~\cite{Zhao_2012_SDUVEEHG_NP} have been successfully preformed at SDUV-FEL, the coherent signal from two-staged cascaded-HGHG and wide frequency tuning of HGHG also have been achieved~\cite{Liu_2013_tunable_PRSTAB}. Now the proof-of-principle of FEL polarization control experiment is ongoing~\cite{Zhang_2012_SDUVpolar_NIMA,Deng_2012_SDUVpolar_FEL2012}. One of the next plan of SDUV-FEL is the high harmonic EEHG lasing, i.e. EEHG-10.
\end{multicols}
\ruleup
\begin{center}
	\includegraphics[width=0.9\textwidth]{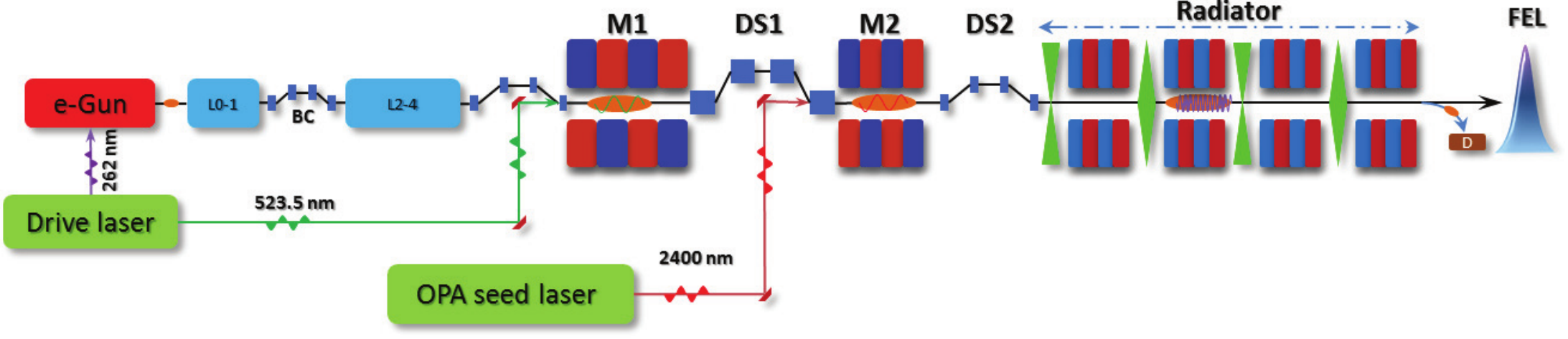}
	\figcaption{ \label{fig:EEHG_layout} Schematic layout of EEHG at SDUV-FEL.}
\end{center}
\ruledown \vspace{0.5cm}

\begin{multicols}{2}
Fig.~\ref{fig:EEHG_layout} is the schematic layout of SDUV-FEL with the classical EEHG configuration. Table~\ref{tab:param} shows the nominal parameters of EEHG-10 for two cases, i.e. with the same or different seed lasers. From Table~\ref{tab:param}, one can find that the two seed lasers are quite different, the optical parametrical amplified (OPA) seed laser can produce the seeds with the wavelength ranging from 1600 to 2600 nm and with pulse duration of about 100 fs full width at half maximum (FWHM); the other seed is the second harmonic of the laser with the wavelength of 1047 nm but with 1 ps pulse width (FWHM). In the following sections numerical simulations of both cases will be presented.

\begin{center}
	\tabcaption{ \label{tab:param} EEHG-10 parameters at SDUV-FEL}
	\footnotesize
	\begin{tabular*}{80mm}{l@{\extracolsep{\fill}}lll}
		\toprule
		Parameter         & Symbol & Value & Unit \\
		\hline
		Beam energy       & $E_b$             & $165$         & $\mathrm{MeV}$ \\
		Energy spread     & $\sigma_\gamma$   & $8.25$        & $\mathrm{keV}$ \\
		Norm Emittance    & $\epsilon_n$      & $2$           & $\mathrm{mm}\cdot\mathrm{mrad}$ \\
		Peak current      & $I_p$             & $100$         & $\mathrm{A}$ \\
		Seed-1 wavelength & $\lambda_s^1$     & $2400$ or $523.5$ & $\mathrm{nm}$ \\
		Seed-2 wavelength & $\lambda_s^2$     & $2400$        & $\mathrm{nm}$ \\		
		Period of Mod-1   & $\lambda_m^1$     & $50$ or $65$  & $\mathrm{mm}$ \\		
		Period of Mod-2   & $\lambda_m^2$     & $50$          & $\mathrm{mm}$ \\		
		Chicane-1         & $R_{56}^1$        & $0-70$        & $\mathrm{mm}$ \\		
		Chicane-2         & $R_{56}^2$        & $0-3.5$       & $\mathrm{mm}$ \\		
		Radiator period   & $\lambda_u$       & $25$          & $\mathrm{mm}$ \\
		FEL wavelength    & $\lambda_\mathrm{FEL}$    & $240$ & $\mathrm{nm}$ \\
		\bottomrule
	\end{tabular*}
\end{center}	

\section{EEHG-10 seeded with two same frequency lasers}
As ref~\cite{Zhao_2012_SDUVEEHG_NP} reported that SDUV-FEL has been lasing at third harmonic of 1047 nm, i.e. 350 nm at the beam energy of 135.4 MeV. The LINAC of SDUV-FEL is supposed to be working at 180 MeV after the upgrade is finished, thus with the proper adjusting of the second seed laser of EEHG, the up-frequency conversion harmonic number could be much higher than 3. In this section, we choose the beam energy of 165 MeV, two seed lasers with the wavelength of 2400 nm which are generated from the OPA system~\cite{Liu_2013_tunable_PRSTAB}. And the harmonic number of EEHG is chosen to be 10, thus the final wavelength of FEL radiation is 240 nm. It is worth to note that the OPA could still be working stably and efficiently at 2400 nm regime while the efficiency decreases rapidly in the longer spectral regime.

In the former configuration of EEHG at SDUV-FEL the period of the first modulator is 65 mm which cannot resonant with the seed of 2400 nm. Luckily, we can use the newly built modulator with the period of 50 mm as the first modulator. However, the new modulator is designed for the FEL polarization control experiment~\cite{Zhang_2012_SDUVpolar_NIMA}, its magnetic field orientation is crossed with conventional version, thus we should tuning the polarized state of the first seed laser to make the interaction between the laser and electron beam happen. The probable hardware arrangement of EEHG-10 with this approach then could be shown as Figure~\ref{fig:case1_layout}.
\end{multicols}
\ruleup
\begin{center}
	\includegraphics[width=0.9\textwidth]{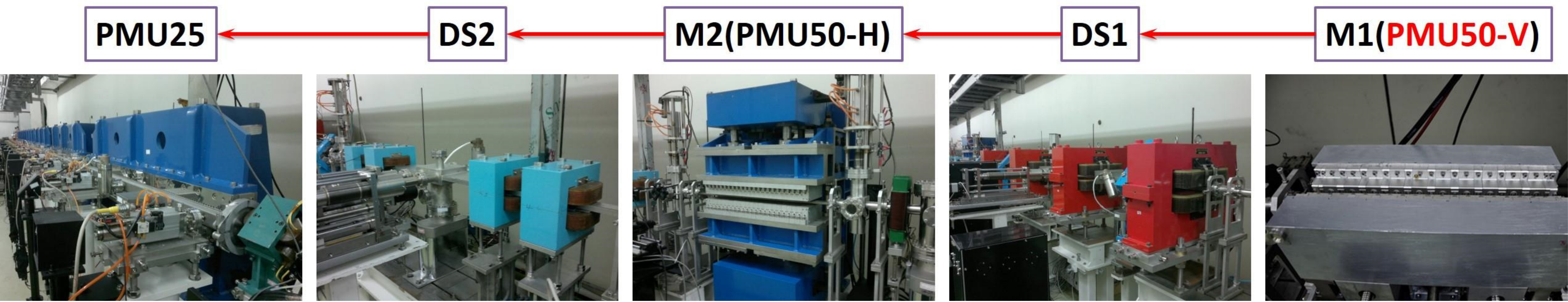}
	\figcaption{ \label{fig:case1_layout} EEHG-10 layout with the same seed lasers, M1, shown as PMU50-V is the newly built vertically polarized planar undulator for FEL polarization control @ SDUV-FEL.}
\end{center}
\ruledown \vspace{0.5cm}
\begin{multicols}{2}

%% phasespace
\begin{center}
	\includegraphics[width=3.9cm]{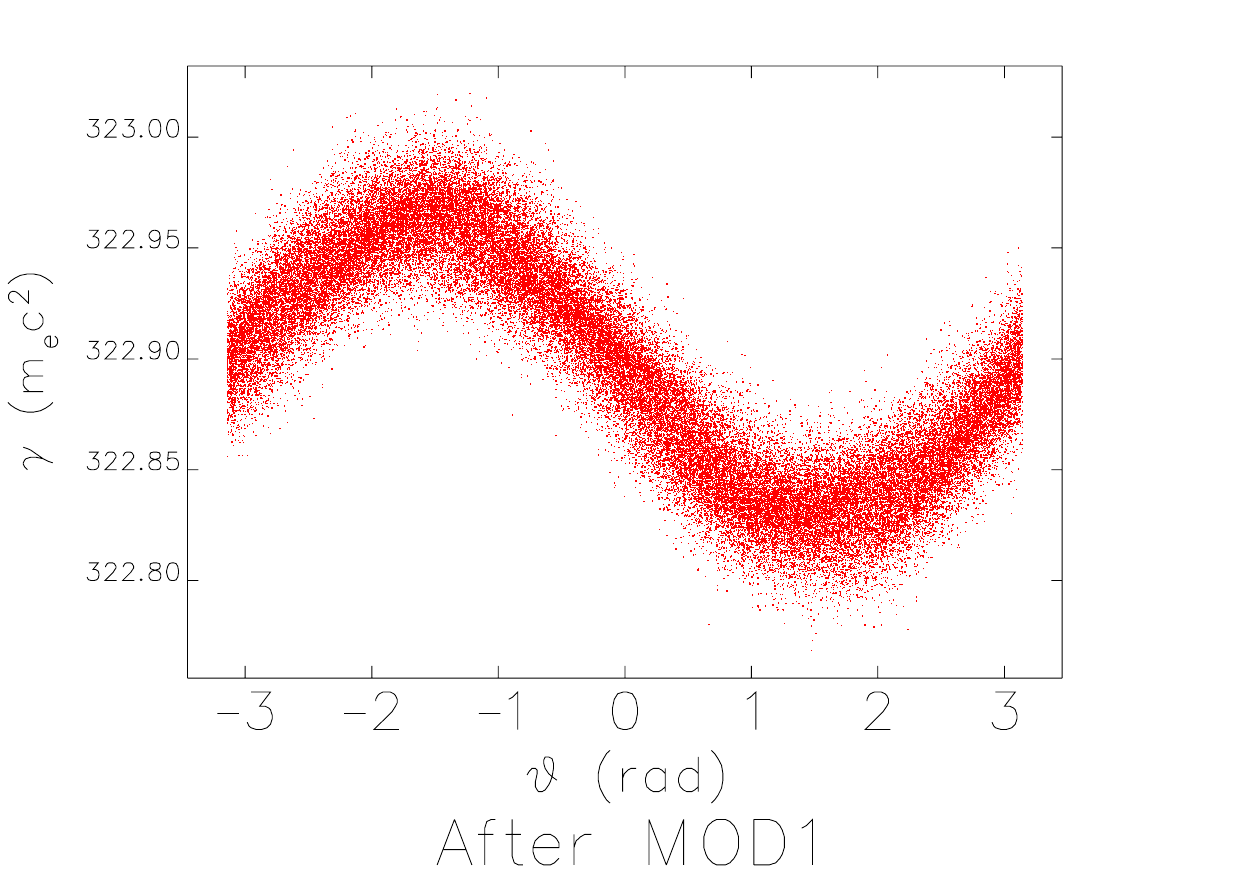}
	\includegraphics[width=3.9cm]{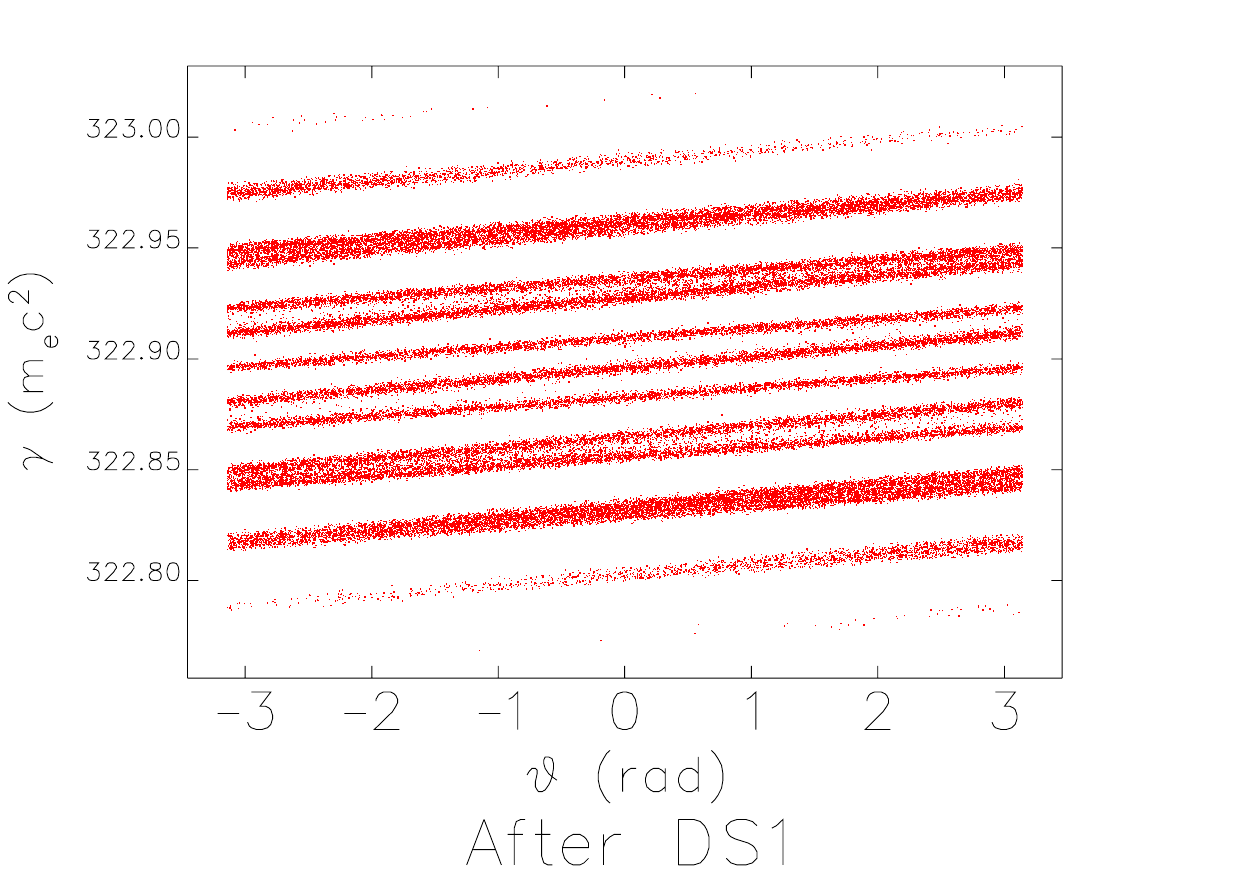}

	\includegraphics[width=3.9cm]{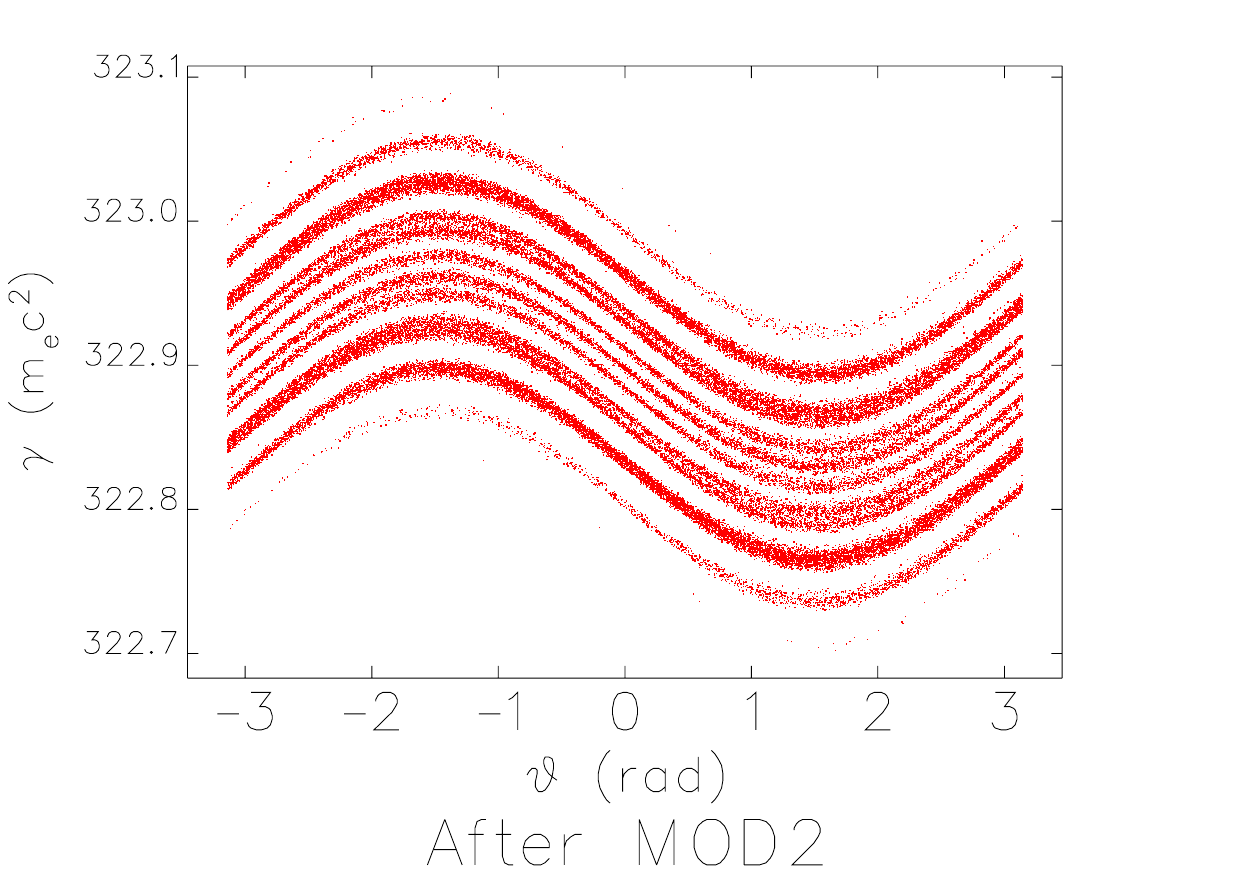}
	\includegraphics[width=3.9cm]{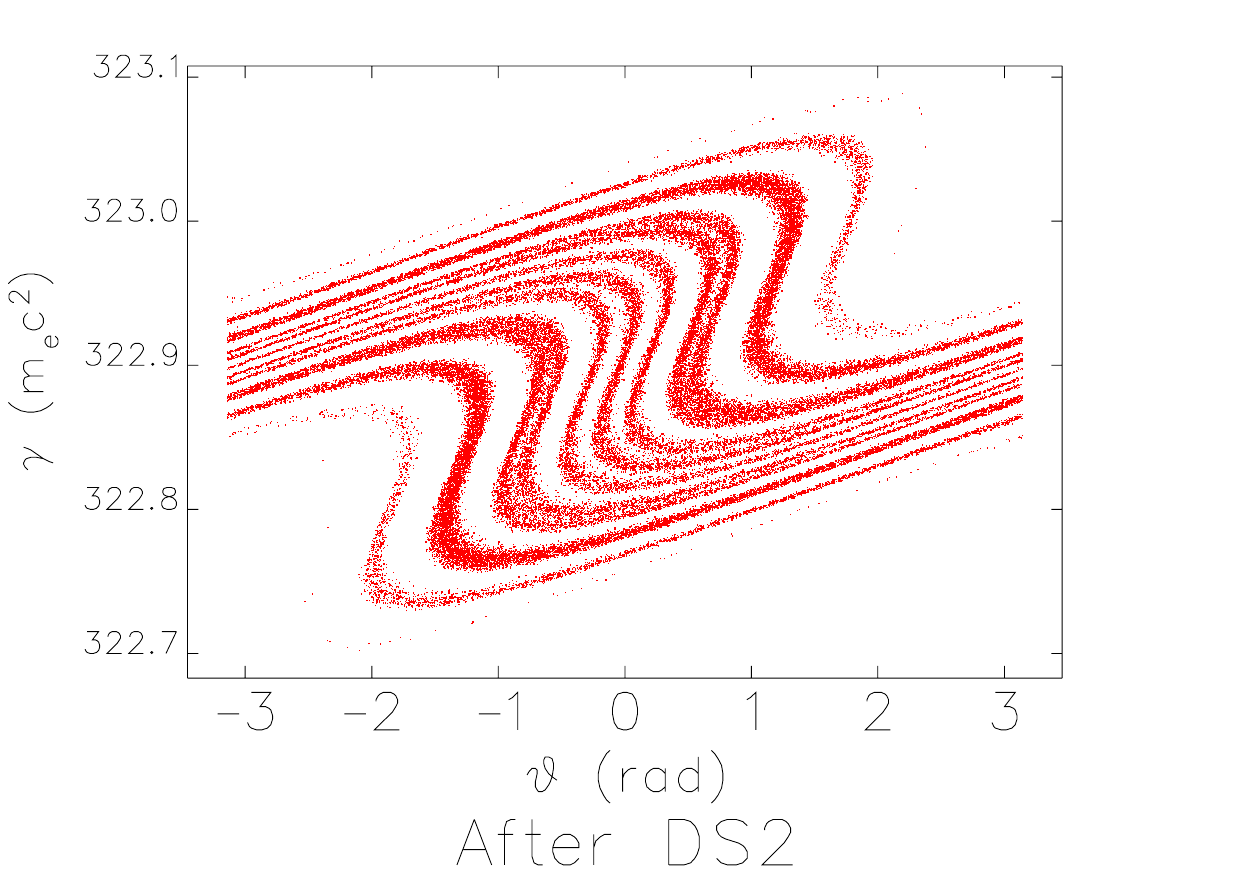}	
	\figcaption{ \label{fig:phasespace} Longitudinal phase space evolution of EEHG-10 at optimal parameters.}
\end{center}

The optimal parameters of EEHG could be found analytically according to the equation of the bunching factor calculation Eqn~\ref{eqn:eqn1}, in which $A_1$ and $A_2$ is the reduced modulation amplitudes in M1 and M2 with the values of $A_i = \Delta\gamma_i/\sigma_\gamma,i=1,2$,$\Delta\gamma_i$ is the modulation amplitude and $\sigma_\gamma$ is the rms local beam energy spread, $B_1$ and $B_2$ is the reduced dispersive strengths in DS1 and DS2 with the values of $B_i = R_{56}^i k_1 \sigma_\gamma/\gamma_0,i = 1,2$, $R_{56}^{1,2}$ is the dispersion of the two chicane, respectively, $k_1$ is the wavenumber of the first seed laser, $gamma_0$ is the central beam energy, $n$ is the up-frequency conversion number, here we choose $n=10$.

\begin{equation}
	\label{eqn:eqn1}
	\scriptsize
	b_n = \left| e^{-\frac{1}{2} \left(-B_1 + n B_2\right){}^2}J_{n+1}\left(-n A_2 B_2\right)J_1\left(A_1\left(B_1-n B_2\right)\right)\right|
\end{equation}

The optimized parameters of EEHG are found as $A_1=A_2=3$, $B_1=3.74$, $B_2=0.42$. Then with the three-dimensional FEL simulation code \textsc{Genesis 1.3}~\cite{Reiche_1999_genesis_NIMA}, the longitudinal phase space evolution of EEHG-10 (see fig~\ref{fig:phasespace}) and the FEL radiation properties have been simulated.

Figure~\ref{fig:gaincurve} shows the gain curve of the average FEL power along the undulator with this approach.

%% average power gain curve
\begin{center}
	\includegraphics[width=8cm]{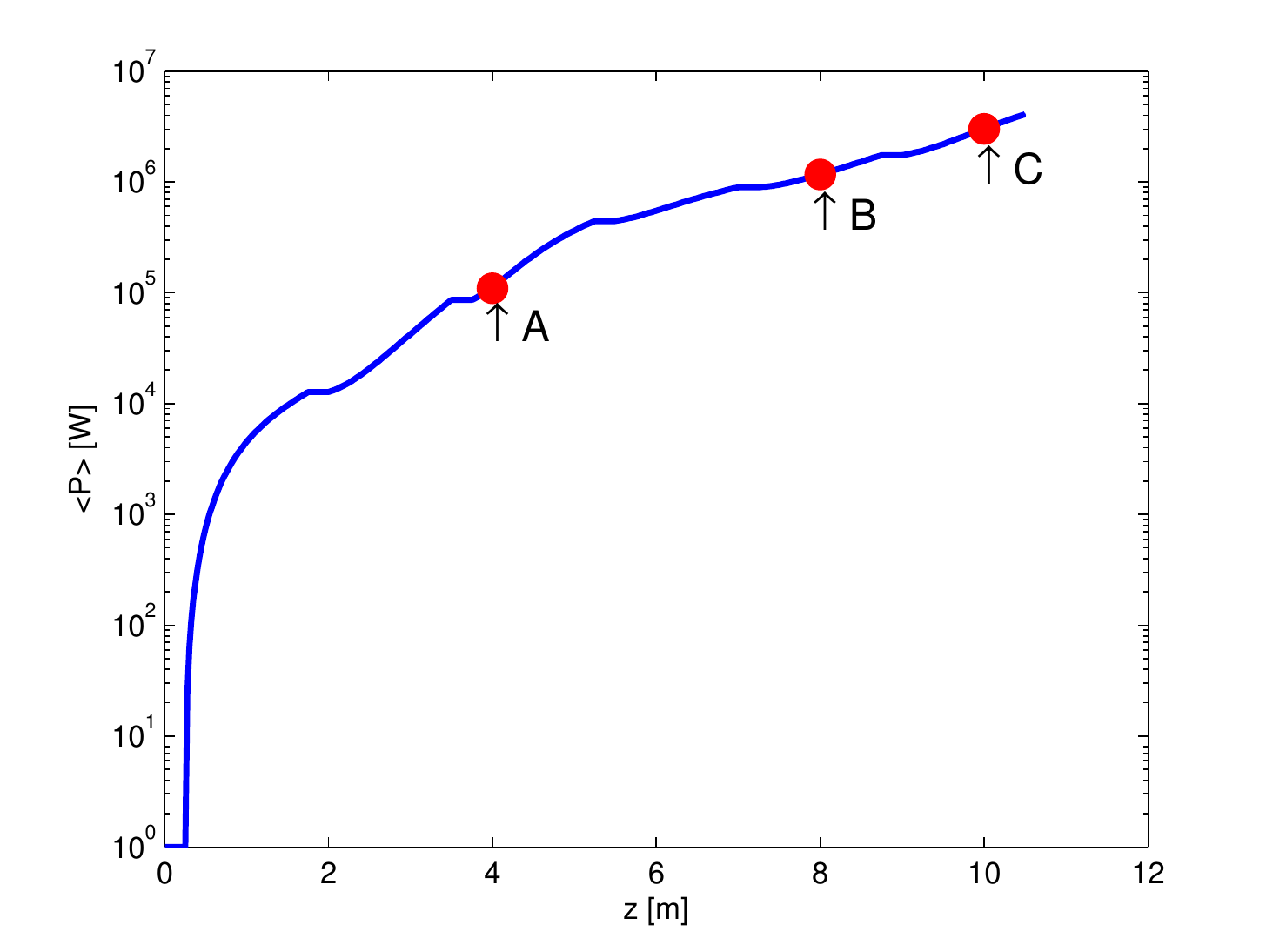}	
	\figcaption{\label{fig:gaincurve} Gain curve of the FEL average power. }
\end{center}

The FEL pulse received at the exit of the last segment of the undulator line can be found from Figure~\ref{fig:FELpulses} and the pulse energy is about $5\,\mathrm{\mu J}$, the spectrum shows the central wavelength is around $240\,\mathrm{nm}$, with almost full longitudinal coherency. One should also notes that the cooperation length of EEHG here is about $200\lambda_s$ which is greater than the FEL pulse length ($125\lambda_s$), thus the superradiance effect contributes the spiky in the frequency domain~\cite{Bonifacio_1990_superradiance_RDNC}. It evidently shows that at the three different positions, A, B and C of the gain curve, the side spikes grow as the Figure~\ref{fig:FELpulses} shows.
\end{multicols}
\ruleup
\begin{center}
	\includegraphics[width=0.3\textwidth]{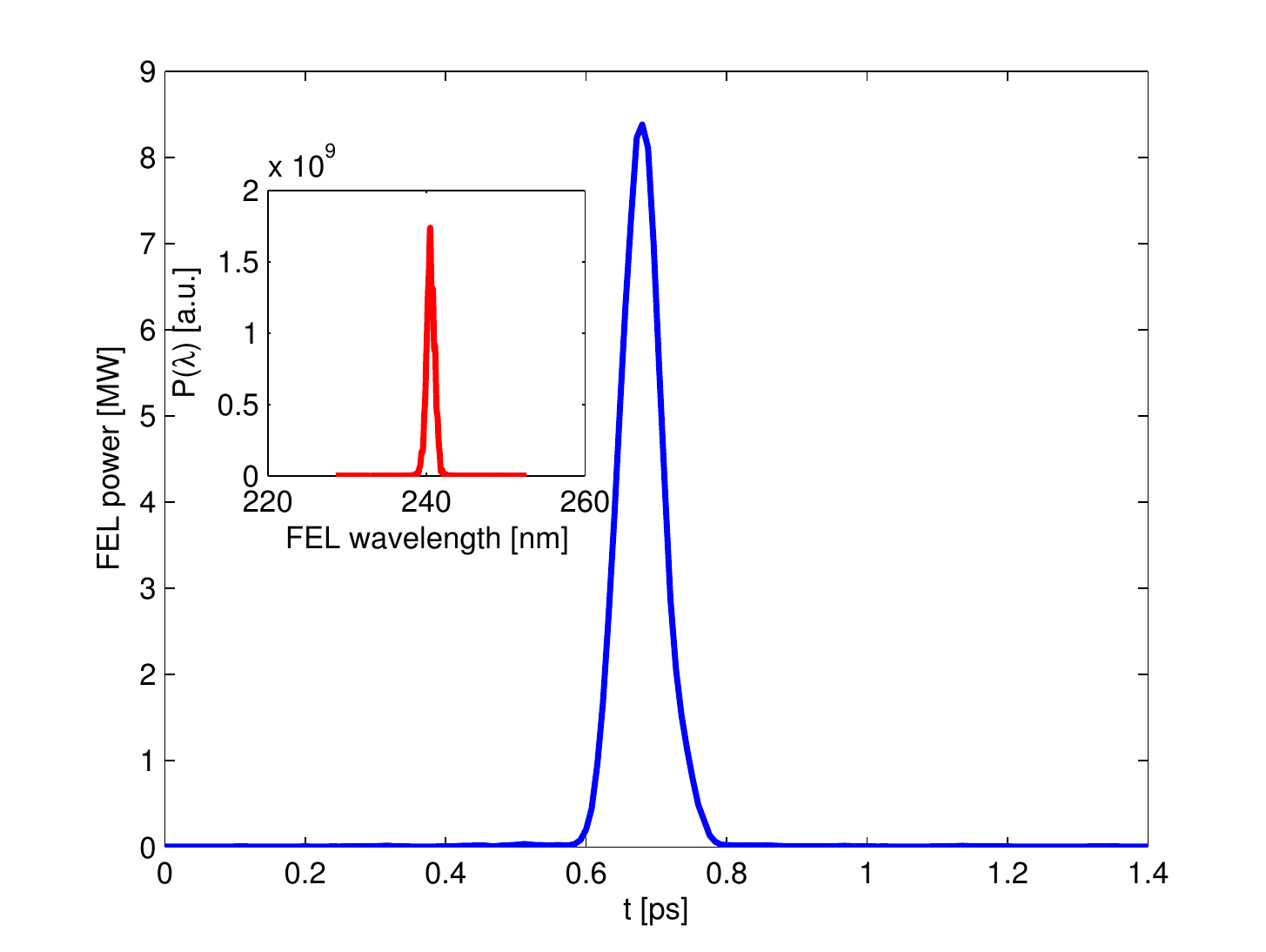}
	\includegraphics[width=0.3\textwidth]{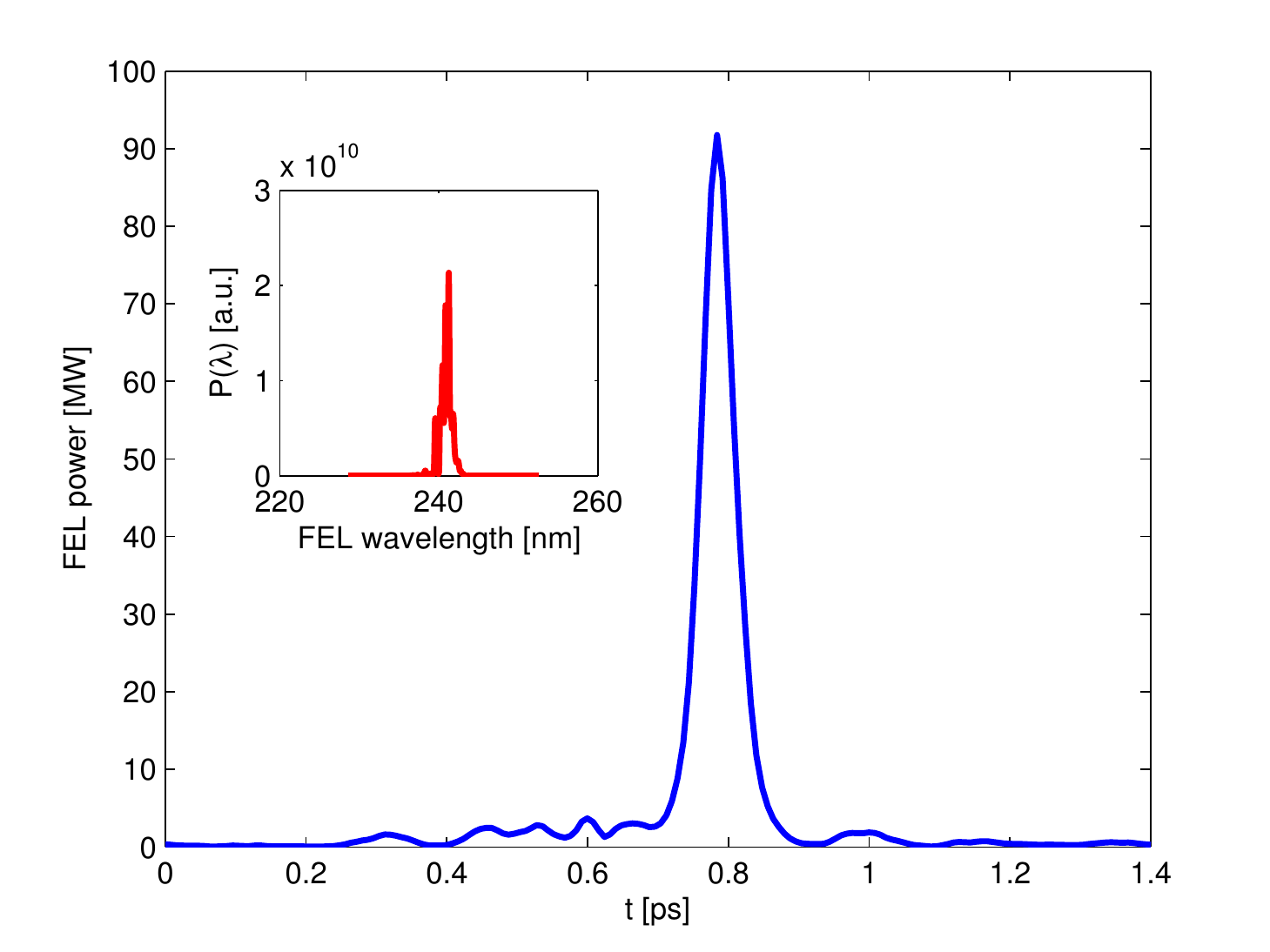}
	\includegraphics[width=0.3\textwidth]{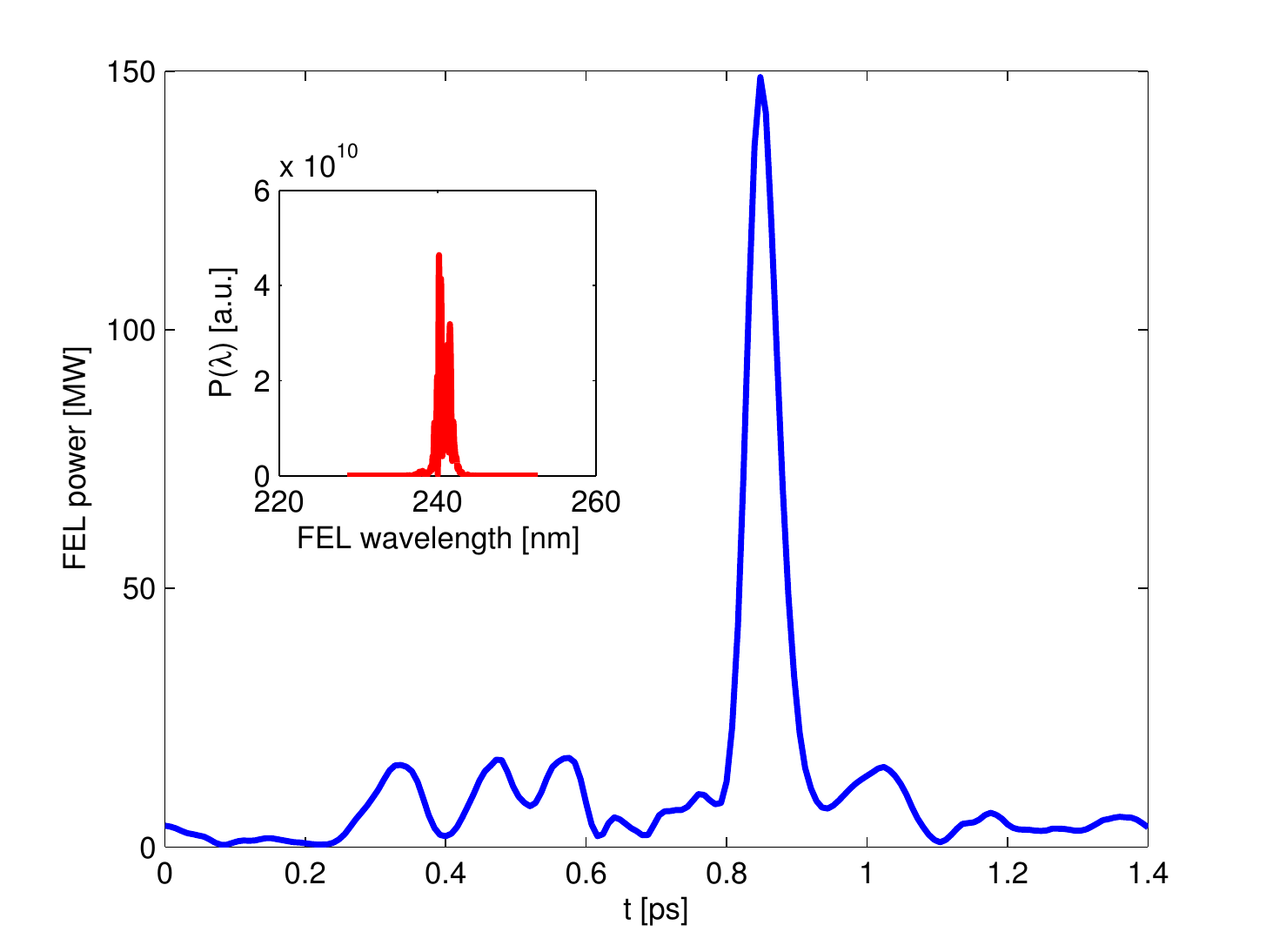}	
	\figcaption{ \label{fig:FELpulses} FEL pulse profile with spectra shown in the subfigure, Left: point A in gaincurve; Middle: point B in gaincurve; Right: point C in gaincurve.}
\end{center}
\ruledown \vspace{0.5cm}
\begin{multicols}{2}

\section{EEHG-10 seeded with two color lasers}
The greatest difference of EEHG is the introducing of the one more additional modulator-chicane section, which is responsible for the generation of the small energy band strips~\cite{Xiang_2009_EEHG_PRSTAB}. Actually, the harmonic number of EEHG is determined by the ratio of frequencies of the two seed lasers. Thus one can also choose the lasers with different colors. In the present layout of SDUV-FEL, the first modulator could be kept still if the first seed of EEHG is shorten to 523.5 nm, that means the hardware configuration does not need to be modified.

The final bunching factor at the $n^{th}$ harmonic of the second seed laser could be formulated as:

\begin{equation}
	\label{eqn:eqn2}
	\scriptsize
	b_n=\left|e^{-\frac{1}{2}\left(-B_1+n\kappa  B_2\right){}^2}J_{n+1/\kappa }\left(-n \kappa  A_2B_2\right)J_1\left(A_1\left(B_1-n \kappa  B_2\right)\right)\right|
\end{equation}

Where $\kappa$ is frequency ratio of the two seed laser, i.e. $\kappa=k_2/k_1$. The optimal conditions can also be figured out analytically,
$A_1=A_2=3$, $B_1=4.997$, $B_2=2.535$.

With this configuration, the layout of EEHG-10 is shown in Figure~\ref{fig:case2_layout}.
\end{multicols}
\ruleup
\begin{center}
	\includegraphics[width=0.9\textwidth]{EEHG10_layout_1.pdf}
	\figcaption{ \label{fig:case2_layout} EEHG-10 layout with the different seed lasers.}
\end{center}
\ruledown \vspace{0.5cm}
\begin{multicols}{2}

The seeds power used for simulation are about 5~MW and 2~MW, respectively. Since the first seed laser is much longer than the second in time domain, the final EEHG signal could be not so much sensitive to the timing jitter effect of the seeds.

The numerical simulation can be divided into two parts. The first is the simulation of the laser beam interaction in the modulator and the other is the FEL simulation in the radiator. We developed efficient code to simulate the laser beam interaction in the modulator with the theory published in ref~\cite{Deng_2010_Laser_NIMA}. The beam dynamics in the dispersive section is tracked by \textsc{Elegant}~\cite{Borland_2000_elegant_LS287}. The simulation of FEL radiation is performed by \textsc{Genesis 1.3} in a self-consistent manner~\cite{Yan_2010_simulation_NIMA}. The whole process is linked by a meticulously designed Makefile from which well-organized scripts are systematically called.

The dispersive strengths optimization result can be found Figure~\ref{fig:EEHG_scan}, from which one can note the best work point of EEHG-10, the left bottom area shows the optimal HGHG work point, however in the EEHG working point, the bunching factor of HGHG is less than 1\% which means the HGHG signal is deeply depressed.

%% scan R56
\begin{center}
	\includegraphics[width=8cm]{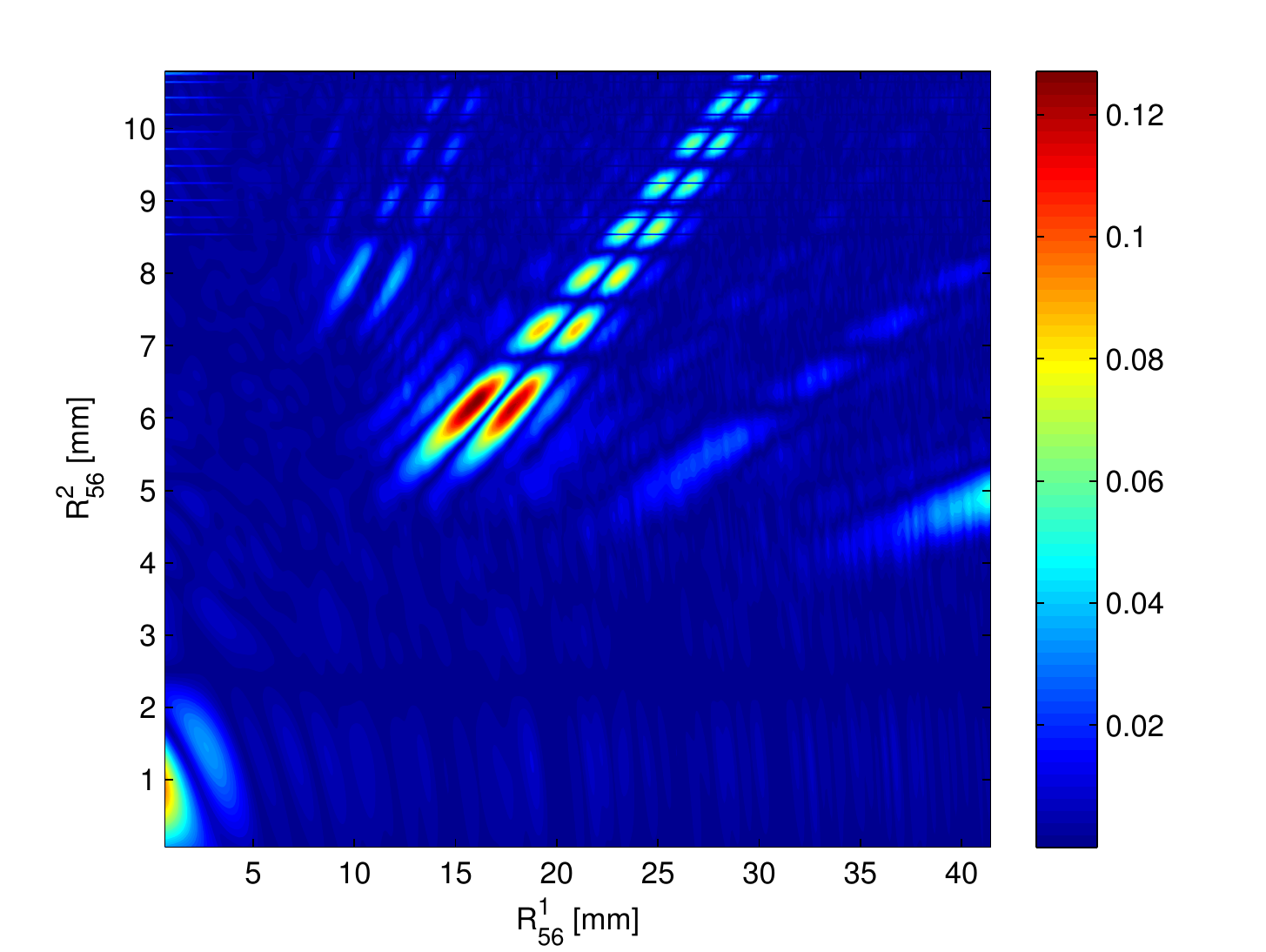}
	\figcaption{ \label{fig:EEHG_scan} Density view of bunching factor at 10th harmonic with the two dispersive strengths.}
\end{center}

It is worth to stress that since the frequency of EEHG-10 is determined by the two seedlaser, the optimal radiation wavelength is not exactly 240 nm. In fact the optimal harmonic number is about 10.4, one can slightly tuning the beam energy to shift the condition.

The coherent signal of EEHG-10 in this case could be detected and verified by the harmonic radiation of ADC undulator (UADC), which is used as the radiator of the first stage of two-staged cascaded-HGHG at SDUV-FEL~\cite{Feng_2012_CascadedHGHG_CSB}. Tuning the magnetic field of UADC, when the resonant wavelength is approaching 690 nm, i.e. three times the wavelength of EEHG-10 signal (230 nm), the bunching factor of third harmonic of 690 nm could be much larger than the fundamental, the simulated gain curve in Figure~\ref{fig:ADC_harm} evidently shows the dominance of the coherent signal from EEHG-10.

\begin{center}
	\includegraphics[width=3.9cm]{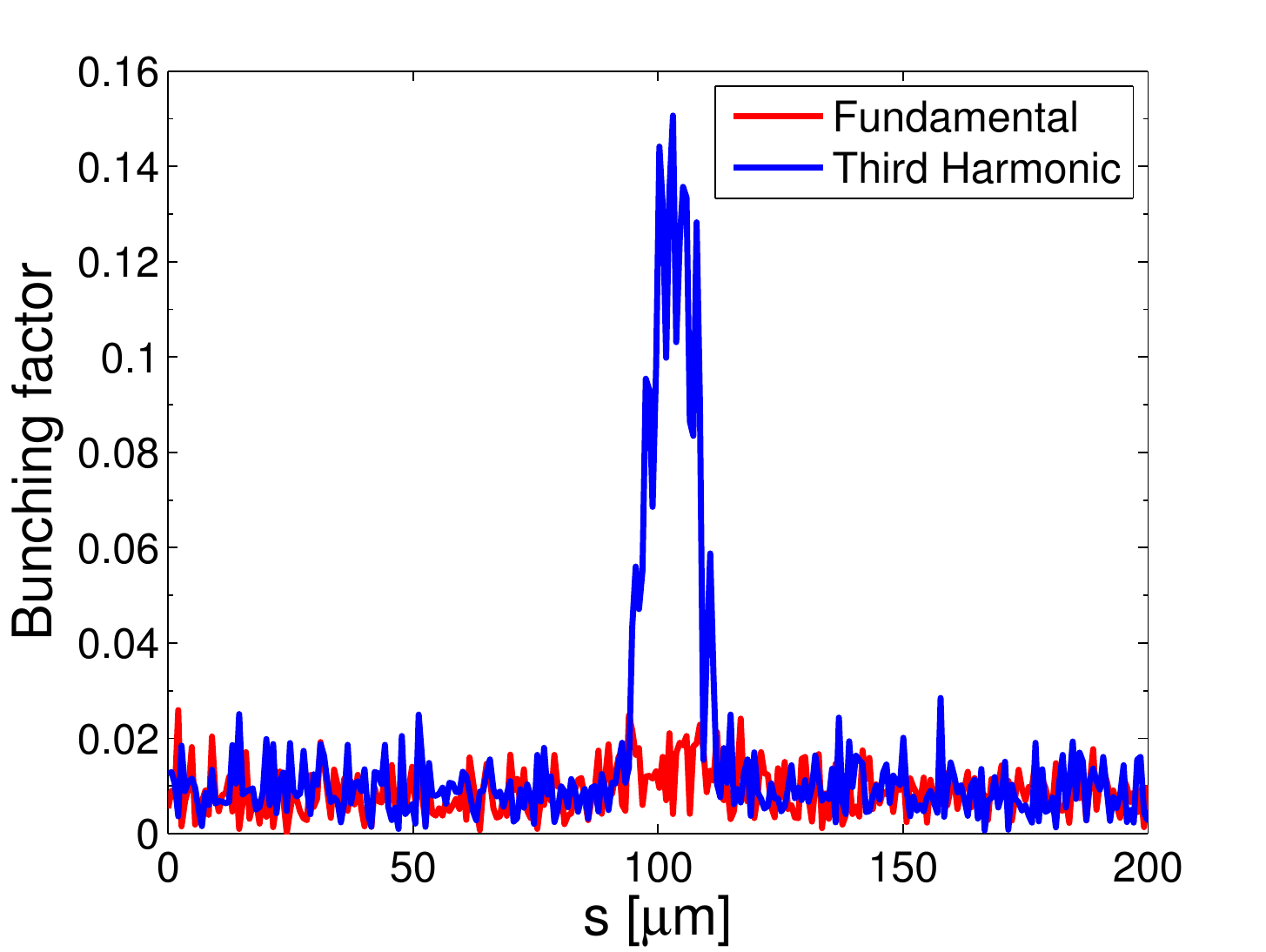}
	\includegraphics[width=3.9cm]{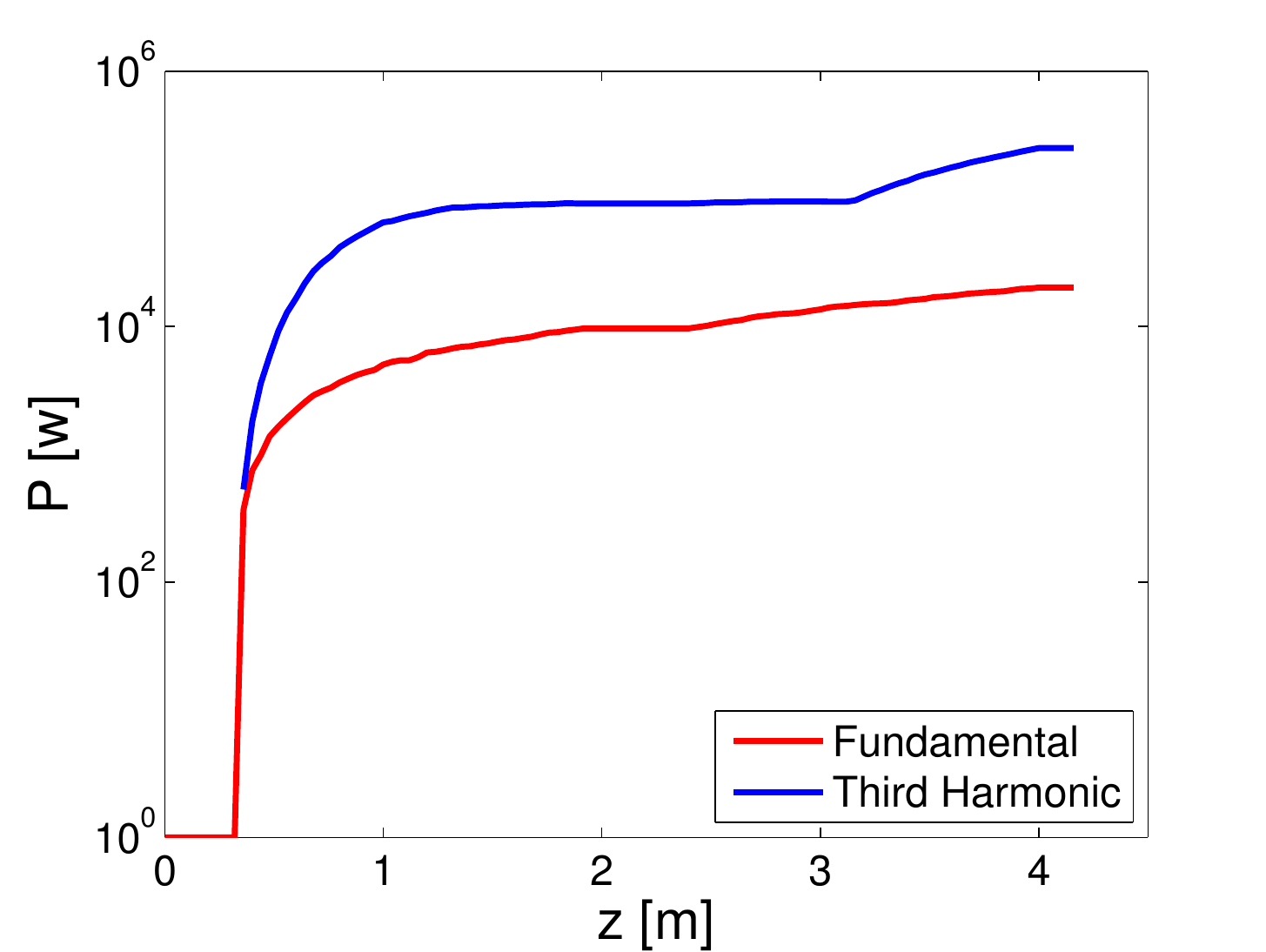}	
	\figcaption{ \label{fig:ADC_harm} Bunching factor and gaincurve for coherent signal diagnostics at EEHG-10.}
\end{center}

%% power and spectra density view
\begin{center}
	\includegraphics[width=8cm]{EEHGavgPower.pdf}
	\figcaption{ \label{fig:EEHG_gc} Gain curve of average FEL power along the undulator.}
\end{center}

After amplified in the long radiator, the FEL power and spectrum could be found from Figure~\ref{fig:EEHG_pulse}. Figure~\ref{fig:EEHG_gc} shows the average FEL power gain curve along the undulator. The simulated FEL pulse energy is about $6\,\mathrm{\mu J}$ at the end of undulator line.

The coherent synchrotron radiation (CSR) effect has also been investigated by the LINAC tracking code---\textsc{Elegant}. The simulation results indicate that CSR induced energy spread will shift the central frequency of the final EEHG-FEL signal, as shown in Figure~\ref{fig:EEHG_pulse}, however the FEL power and the gain curve are not much affected since the sliced parameters of EEHG almost keep the same level.

%% average gain curve and power profile
\begin{center}
	\includegraphics[width=3.95cm]{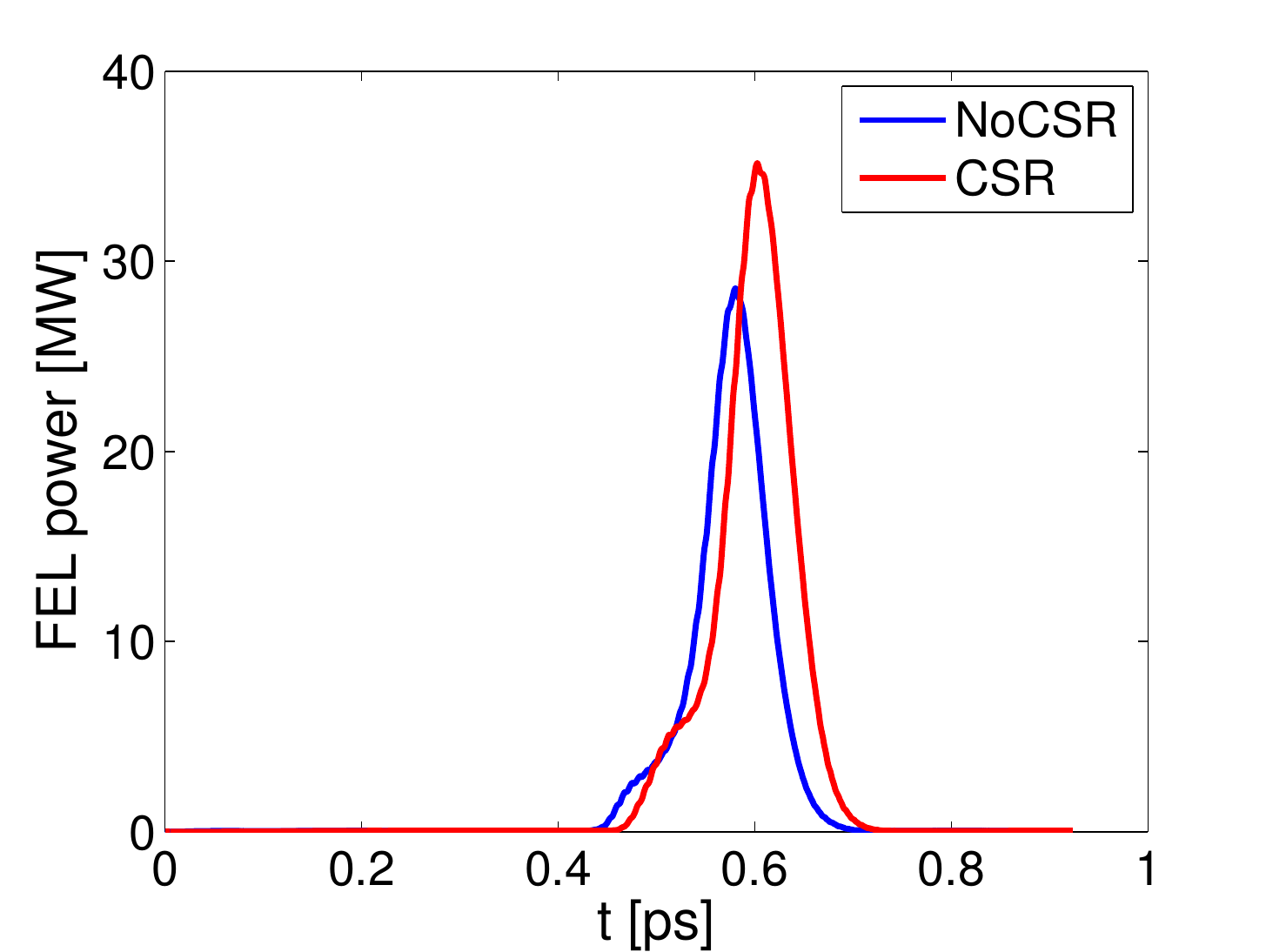}
	\includegraphics[width=3.95cm]{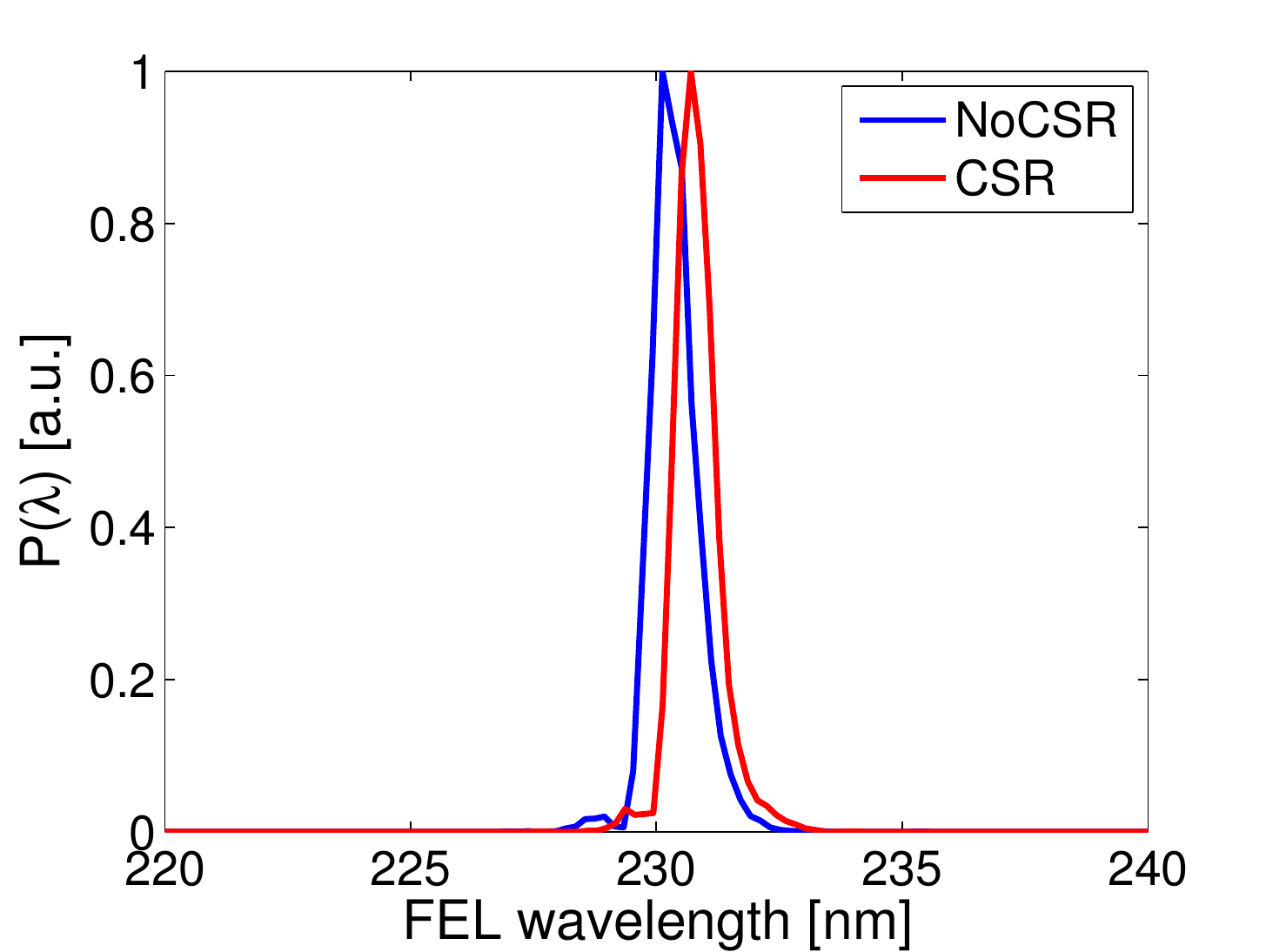}	
	\figcaption{ \label{fig:EEHG_pulse} Typical EEHG pulse profile in time (left) and frequency (right) domain. }
\end{center}

\section{Conclusions}
With the successful demonstration and the first lasing of EEHG at third harmonic, it becomes more and more significant to make the EEHG lasing at much higher harmonics so as to pave the way to even higher harmonic lasing, e.g. generating the fully coherent hard X-ray FELs in a single EEHG stage~\cite{Zhao_2012_interview_NP}. Lasing at 10-th harmonic of the second seed laser of EEHG at SDUV-FEL has been proposed, two different experimental plans have been carefully studied. Full three dimensional numerical simulations indicate that both approaches could lead to the lasing of EEHG at 10-th harmonic of the seed laser. The two-color seeded EEHG-10 could be performed at SDUV-FEL with the present hardware configuration but need much more sophisticated seed optical configuration, while seeding with the same seed source requires the polarization tuning of the seed laser. However the numerical simulation results show a clear insight into the lasing of EEHG-10 at SDUV-FEL, which, of course will illuminate the way of the EEHG at SXFEL into the X-ray regime.

\vspace{0.5cm}

\acknowledgments{The authors would like to thank WANG Guang-Lei, YAO Hai-Feng, WANG Xing-Tao, CHEN Jian-Hui, SHEN Lei, LAN Tai-He and LIU Bo for help discussions.}

\vspace{0.5cm}

% references
\bibliographystyle{model1a-num-names}
\bibliography{refs}

\begin{thebibliography}{36}
\expandafter\ifx\csname natexlab\endcsname\relax\def\natexlab#1{#1}\fi
\providecommand{\bibinfo}[2]{#2}
\ifx\xfnm\relax \def\xfnm[#1]{\unskip,\space#1}\fi
%Type = Article
\bibitem[{Madey(1971)}]{Madey_1971_FEL_JAP}
\bibinfo{author}{J.~M.~J. Madey}, \bibinfo{journal}{Journal of Applied Physics}
  \bibinfo{volume}{42} (\bibinfo{year}{1971}) \bibinfo{pages}{1906}.
%Type = Article
\bibitem[{O'Shea and Freund(2001)}]{OShea_2001_FEL_Science}
\bibinfo{author}{P.~G. O'Shea}, \bibinfo{author}{H.~P. Freund},
  \bibinfo{journal}{Science} \bibinfo{volume}{292} (\bibinfo{year}{2001})
  \bibinfo{pages}{1853}.
%Type = Article
\bibitem[{Barletta et~al.(2010)Barletta, Bisognano, Corlett
  et~al.}]{Barletta_2010_FEL_NIMA}
\bibinfo{author}{W.~A. Barletta}, \bibinfo{author}{J.~Bisognano},
  \bibinfo{author}{J.~N. Corlett}, et~al., \bibinfo{journal}{Nuclear
  Instruments and Methods in Physics Research Section A: Accelerators,
  Spectrometers, Detectors and Associated Equipment} \bibinfo{volume}{618}
  (\bibinfo{year}{2010}) \bibinfo{pages}{69--96}.
%Type = Article
\bibitem[{McNeil and Thompson(2010)}]{McNeil_2010_XFEL_NP}
\bibinfo{author}{B.~W.~J. McNeil}, \bibinfo{author}{N.~R. Thompson},
  \bibinfo{journal}{Nature Photonics} \bibinfo{volume}{4}
  (\bibinfo{year}{2010}) \bibinfo{pages}{814--821}.
%Type = Article
\bibitem[{Emma et~al.(2010)Emma, Akre, Arthur et~al.}]{Emma_2010_LCLS_NP}
\bibinfo{author}{P.~Emma}, \bibinfo{author}{R.~Akre},
  \bibinfo{author}{J.~Arthur}, et~al., \bibinfo{journal}{Nature Photonics}
  \bibinfo{volume}{4} (\bibinfo{year}{2010}) \bibinfo{pages}{641--647}.
%Type = Article
\bibitem[{Ishikawa et~al.(2012)Ishikawa, Aoyagi, Asaka
  et~al.}]{Ishikawa_2012_SACLA_NP}
\bibinfo{author}{T.~Ishikawa}, \bibinfo{author}{H.~Aoyagi},
  \bibinfo{author}{T.~Asaka}, et~al., \bibinfo{journal}{Nature Photonics}
  \bibinfo{volume}{6} (\bibinfo{year}{2012}) \bibinfo{pages}{540--544}.
%Type = Techreport
\bibitem[{Altarelli et~al.(2007)Altarelli, Brinkmann, Chergui
  et~al.}]{Massimo_2007_CDREXFEL}
\bibinfo{author}{M.~Altarelli}, \bibinfo{author}{R.~Brinkmann},
  \bibinfo{author}{M.~Chergui}, et~al., \bibinfo{title}{The European X-Ray
  Free-Electron Laser Technical design report}, \bibinfo{type}{Technical
  Report}, \bibinfo{year}{2007}.
%Type = Article
\bibitem[{Altarelli(2011)}]{Altarelli_2011_EXFEL_NIMB}
\bibinfo{author}{M.~Altarelli}, \bibinfo{journal}{Nuclear Instruments and
  Methods in Physics Research Section B: Beam Interactions with Materials and
  Atoms} \bibinfo{volume}{269} (\bibinfo{year}{2011})
  \bibinfo{pages}{2845--2849}.
%Type = Misc
\bibitem[{PSI(2010)}]{SwissFEL_CDR}
\bibinfo{author}{PSI}, \bibinfo{title}{Swissfel conceptual design report},
  \bibinfo{year}{2010}.
%Type = Article
\bibitem[{Bonifacio et~al.(1984)Bonifacio, Pellegrini, and
  Narducci}]{Bonifacio_1984_FEL_OC}
\bibinfo{author}{R.~Bonifacio}, \bibinfo{author}{C.~Pellegrini},
  \bibinfo{author}{L.~M. Narducci}, \bibinfo{journal}{Optics Communications}
  \bibinfo{volume}{50} (\bibinfo{year}{1984}) \bibinfo{pages}{373--378}.
%Type = Article
\bibitem[{Milton et~al.(2001)Milton, Gluskin, Arnold
  et~al.}]{Milton_2001_SASE_Science}
\bibinfo{author}{S.~V. Milton}, \bibinfo{author}{E.~Gluskin},
  \bibinfo{author}{N.~D. Arnold}, et~al., \bibinfo{journal}{Science}
  \bibinfo{volume}{292} (\bibinfo{year}{2001}) \bibinfo{pages}{2037}.
%Type = Article
\bibitem[{Yu et~al.(2000)Yu, Babzien, Ben-Zvi et~al.}]{Yu_2000_HGHG_Science}
\bibinfo{author}{L.~H. Yu}, \bibinfo{author}{M.~Babzien},
  \bibinfo{author}{I.~Ben-Zvi}, et~al., \bibinfo{journal}{Science}
  \bibinfo{volume}{289} (\bibinfo{year}{2000}) \bibinfo{pages}{932--934}.
%Type = Article
\bibitem[{Jia(2008)}]{Jia_2008_EHGHG_APL}
\bibinfo{author}{Q.~Jia}, \bibinfo{journal}{Applied Physics Letters}
  \bibinfo{volume}{93} (\bibinfo{year}{2008}) \bibinfo{pages}{141102}.
%Type = Article
\bibitem[{Stupakov(2009)}]{Stupakov_2009_EEHG_PRL}
\bibinfo{author}{G.~Stupakov}, \bibinfo{journal}{Physical Review Letters}
  \bibinfo{volume}{102} (\bibinfo{year}{2009}) \bibinfo{pages}{074801}.
%Type = Article
\bibitem[{Xiang et~al.(2010)Xiang, Colby, Dunning et~al.}]{Xiang_2010_EEHG_PRL}
\bibinfo{author}{D.~Xiang}, \bibinfo{author}{E.~Colby},
  \bibinfo{author}{M.~Dunning}, et~al., \bibinfo{journal}{Physical Review
  Letters} \bibinfo{volume}{105} (\bibinfo{year}{2010})
  \bibinfo{pages}{114801}.
%Type = Article
\bibitem[{Zhao et~al.(2012)Zhao, Wang, Chen et~al.}]{Zhao_2012_SDUVEEHG_NP}
\bibinfo{author}{Z.~T. Zhao}, \bibinfo{author}{D.~Wang}, \bibinfo{author}{J.~H.
  Chen}, et~al., \bibinfo{journal}{Nature Photonics} \bibinfo{volume}{6}
  (\bibinfo{year}{2012}) \bibinfo{pages}{360--363}.
%Type = Article
\bibitem[{Amann et~al.(2012)Amann, Berg, Blank
  et~al.}]{Amann_2012_LCLSself-seeding_NP}
\bibinfo{author}{J.~Amann}, \bibinfo{author}{W.~Berg},
  \bibinfo{author}{V.~Blank}, et~al., \bibinfo{journal}{Nat Photon}
  \bibinfo{volume}{6} (\bibinfo{year}{2012}) \bibinfo{pages}{693--698}.
%Type = Article
\bibitem[{Allaria et~al.(2012)Allaria, Appio, Badano
  et~al.}]{Allaria_2012_FermiFEL_NP}
\bibinfo{author}{E.~Allaria}, \bibinfo{author}{R.~Appio},
  \bibinfo{author}{L.~Badano}, et~al., \bibinfo{journal}{Nature Photonics}
  \bibinfo{volume}{6} (\bibinfo{year}{2012}) \bibinfo{pages}{699--704}.
%Type = Article
\bibitem[{Doyuran et~al.(2001)Doyuran, Babzien, Shaftan
  et~al.}]{Doyuran_2001_BNL_PRL}
\bibinfo{author}{A.~Doyuran}, \bibinfo{author}{M.~Babzien},
  \bibinfo{author}{T.~Shaftan}, et~al., \bibinfo{journal}{Physical Review
  Letters} \bibinfo{volume}{86} (\bibinfo{year}{2001})
  \bibinfo{pages}{5902--5905}.
%Type = Article
\bibitem[{Zhao et~al.(2004)Zhao, Dai, Zhao et~al.}]{Zhao_2004_SDUV_NIMA}
\bibinfo{author}{Z.~Zhao}, \bibinfo{author}{Z.~Dai}, \bibinfo{author}{X.~Zhao},
  et~al., \bibinfo{journal}{Nuclear Instruments and Methods in Physics Research
  Section A: Accelerators, Spectrometers, Detectors and Associated Equipment}
  \bibinfo{volume}{528} (\bibinfo{year}{2004}) \bibinfo{pages}{591--594}.
%Type = Inproceedings
\bibitem[{Zhao and Wang(2010)}]{Zhao_2010_SDUV-FEL_FEL10}
\bibinfo{author}{Z.~Zhao}, \bibinfo{author}{D.~Wang}, in:
  \bibinfo{booktitle}{FEL10}.
%Type = Article
\bibitem[{Feldhaus(2010)}]{Feldhaus_2010_FLASH_JPB}
\bibinfo{author}{J.~Feldhaus}, \bibinfo{journal}{Journal of Physics B: Atomic,
  Molecular and Optical Physics} \bibinfo{volume}{43} (\bibinfo{year}{2010})
  \bibinfo{pages}{194002}.
%Type = Inproceedings
\bibitem[{Corlett et~al.(2011)Corlett, Austin, Baptiste
  et~al.}]{Corlett_2011_NGLS_PAC2011}
\bibinfo{author}{J.~Corlett}, \bibinfo{author}{B.~Austin},
  \bibinfo{author}{K.~Baptiste}, et~al., in: \bibinfo{booktitle}{Particle
  Accelerator Conference}, pp. \bibinfo{pages}{775--777}.
%Type = Techreport
\bibitem[{SINAP et~al.(2008)SINAP, IHEP, and THU}]{SXFEL_CDR}
\bibinfo{author}{SINAP}, \bibinfo{author}{IHEP}, \bibinfo{author}{THU},
  \bibinfo{title}{Shanghai Soft X-ray FEL Concept Design Report},
  \bibinfo{type}{Technical Report}, \bibinfo{year}{2008}.
%Type = Inproceedings
\bibitem[{Li et~al.(2010)Li, Chen, Deng et~al.}]{Dongguo_2010_SDUVSASE_FEL10}
\bibinfo{author}{D.~Li}, \bibinfo{author}{J.~Chen}, \bibinfo{author}{H.~Deng},
  et~al., in: \bibinfo{booktitle}{International Free Electron Laser
  Conference}, p. \bibinfo{pages}{WEPA02}.
%Type = Article
\bibitem[{Liu et~al.(2013)Liu, Li, Chen et~al.}]{Liu_2013_tunable_PRSTAB}
\bibinfo{author}{B.~Liu}, \bibinfo{author}{W.~B. Li}, \bibinfo{author}{J.~H.
  Chen}, et~al., \bibinfo{journal}{Physical Review Special Topics -
  Accelerators and Beams} \bibinfo{volume}{16} (\bibinfo{year}{2013})
  \bibinfo{pages}{020704}. \bibinfo{note}{PRSTAB}.
%Type = Article
\bibitem[{Zhang et~al.(2012)Zhang, Deng, Chen
  et~al.}]{Zhang_2012_SDUVpolar_NIMA}
\bibinfo{author}{T.~Zhang}, \bibinfo{author}{H.~Deng},
  \bibinfo{author}{J.~Chen}, et~al., \bibinfo{journal}{Nuclear Instruments and
  Methods in Physics Research Section A: Accelerators, Spectrometers, Detectors
  and Associated Equipment} \bibinfo{volume}{680} (\bibinfo{year}{2012})
  \bibinfo{pages}{112--116}.
%Type = Inproceedings
\bibitem[{Deng et~al.(2012)Deng, Zhang, Feng
  et~al.}]{Deng_2012_SDUVpolar_FEL2012}
\bibinfo{author}{H.~Deng}, \bibinfo{author}{T.~Zhang},
  \bibinfo{author}{L.~Feng}, et~al., in: \bibinfo{booktitle}{FEL 2012}.
%Type = Article
\bibitem[{Reiche(1999)}]{Reiche_1999_genesis_NIMA}
\bibinfo{author}{S.~Reiche}, \bibinfo{journal}{Nuclear Instruments and Methods
  in Physics Research Section A: Accelerators, Spectrometers, Detectors and
  Associated Equipment} \bibinfo{volume}{429} (\bibinfo{year}{1999})
  \bibinfo{pages}{243--248}.
%Type = Article
\bibitem[{Bonifacio et~al.(1990)Bonifacio, Casagrande, Cerchioni
  et~al.}]{Bonifacio_1990_superradiance_RDNC}
\bibinfo{author}{R.~Bonifacio}, \bibinfo{author}{F.~Casagrande},
  \bibinfo{author}{G.~Cerchioni}, et~al., \bibinfo{journal}{RIVISTA DEL NUOVO
  CIMENTO} \bibinfo{volume}{13} (\bibinfo{year}{1990}) \bibinfo{pages}{1--69}.
%Type = Article
\bibitem[{Xiang and Stupakov(2009)}]{Xiang_2009_EEHG_PRSTAB}
\bibinfo{author}{D.~Xiang}, \bibinfo{author}{G.~Stupakov},
  \bibinfo{journal}{Physical Review Special Topics - Accelerators and Beams}
  \bibinfo{volume}{12} (\bibinfo{year}{2009}) \bibinfo{pages}{030702}.
%Type = Article
\bibitem[{Deng et~al.(2010)Deng, Yan, Wang et~al.}]{Deng_2010_Laser_NIMA}
\bibinfo{author}{H.~X. Deng}, \bibinfo{author}{J.~Yan},
  \bibinfo{author}{D.~Wang}, et~al., \bibinfo{journal}{Nuclear Instruments and
  Methods in Physics Research Section A: Accelerators, Spectrometers, Detectors
  and Associated Equipment} \bibinfo{volume}{622} (\bibinfo{year}{2010})
  \bibinfo{pages}{508--511}.
%Type = Article
\bibitem[{Borland(2000)}]{Borland_2000_elegant_LS287}
\bibinfo{author}{M.~Borland}, \bibinfo{journal}{elegant: A Flexible
  SDDS-Compliant Code for Accelerator Simulation, Advance Photon Source LS-287}
   (\bibinfo{year}{2000}).
%Type = Article
\bibitem[{Yan et~al.(2010)Yan, Zhang, and Deng}]{Yan_2010_simulation_NIMA}
\bibinfo{author}{J.~Yan}, \bibinfo{author}{M.~Zhang}, \bibinfo{author}{H.-X.
  Deng}, \bibinfo{journal}{Nuclear Instruments and Methods in Physical Research
  Section A} \bibinfo{volume}{615} (\bibinfo{year}{2010})
  \bibinfo{pages}{249--253}.
%Type = Article
\bibitem[{Feng et~al.(2012)Feng, Zhang, Lin
  et~al.}]{Feng_2012_CascadedHGHG_CSB}
\bibinfo{author}{C.~Feng}, \bibinfo{author}{M.~Zhang},
  \bibinfo{author}{G.~Lin}, et~al., \bibinfo{journal}{Chinese Science Bulletin}
  \bibinfo{volume}{57} (\bibinfo{year}{2012}) \bibinfo{pages}{3423--3429}.
%Type = Article
\bibitem[{Won and Zhao(2012)}]{Zhao_2012_interview_NP}
\bibinfo{author}{R.~Won}, \bibinfo{author}{Z.~T. Zhao},
  \bibinfo{journal}{Nature Photonics} \bibinfo{volume}{6}
  (\bibinfo{year}{2012}) \bibinfo{pages}{406--406}.

\end{thebibliography}

\end{multicols}

\clearpage

%\end{CJK}

\end{document}